\documentstyle[twoside,fleqn,hyperon99_paper]{article}   
   
\newcommand{\ri}{\mbox{$\rm i$}}   
\newcommand{\bfm}[1]{\mbox{\boldmath$#1$}}  
\newcommand{\ratio}[2]{\mbox{$#1\over#2$}}  
  
\newcommand{\AmS}{{\protect\the\textfont2  
  A\kern-.1667em\lower.5ex\hbox{M}\kern-.125emS}}  
  
\title{%
\hfill \raisebox{2ex}{\rm\small ISU-HET-99-14} \\   
Recent theoretical results on  $|\Delta\bfm{I}|=3/2$  decays of   
hyperons\thanks{%
Talk presented at Hyperon99 Symposium, Fermilab, Batavia, 27-29 
September 1999. 
This work  was supported in part by DOE under contract number 
DE-FG02-92ER40730.}}  
 
\author{Jusak Tandean \address{Department of Physics and Astronomy,\\
        Iowa State University,\\ Ames, IA 50010}}
       
\begin{document}

\begin{abstract}
We present a discussion of the  $\,|\Delta\bfm{I}|=3/2\,$  amplitudes 
of the hyperon decays  $\,B\rightarrow B'\pi\,$  in the context of 
chiral perturbation theory.  
We evaluate the theoretical uncertainty of the lowest-order predictions by 
calculating the leading non-analytic corrections.  
We find that the corrections to the lowest-order predictions are 
within the expectations of naive power-counting and, therefore, 
that this picture can be examined more quantitatively with improved 
measurements.  
\end{abstract}

\maketitle


Hyperon nonleptonic decays have been much studied within 
the framework of chiral perturbation theory~($\chi$PT).  
The decay modes are  $\,\Sigma^+\rightarrow n\pi^+,\,$  
$\,\Sigma^+\rightarrow p\pi^0,\,$  
$\Sigma^-\rightarrow n\pi^-,\,$  
$\Lambda\rightarrow p\pi^-,\,$  $\Lambda\rightarrow n\pi^0,\,$  
$\Xi^-\rightarrow\Lambda\pi^-,\,$ 
and  $\,\Xi^0\rightarrow\Lambda\pi^0.\,$
Most of the calculations have dealt with the dominant  
$\,|\Delta\bfm{I}|=1/2\,$  amplitudes of these decays, and 
the results have been 
mixed~\cite{Bijn85,Geor84,Jenk92a,Jenk92b,Caro92,Spri95,Abde99a}. 
The theory can well reproduce either the S-waves or 
the P-waves, but not both simultaneously.

The  $\,|\Delta\bfm{I}|=3/2\,$  amplitudes of these decays have not 
been well studied in~$\chi$PT.   
In view of the situation in the  $\,|\Delta\bfm{I}|=1/2\,$  
sector, it is, therefore, instructive to carry out a~similar analysis 
in the  $\,|\Delta\bfm{I}|=3/2\,$  sector.  
Such an analysis has recently been done~\cite{Abde99b}, 
and some of its results will be presented here.

To apply~$\chi$PT to interactions involving the lowest-lying mesons 
and baryons, we employ the heavy-baryon 
formalism~\cite{Jenk92b,Jenk91a}.  
In this approach, the theory has a~consistent chiral expansion,   
and the baryons in the effective chiral Lagrangian are described by 
velocity-dependent fields.   
Here, we include both octet and decuplet baryons in the Lagrangian 
because the octet-decuplet mass difference is small enough to make 
the effects of the decuplet significant on the low-energy 
theory~\cite{Jenk92b,Jenk91b}.

The leading-order chiral Lagrangian for the strong interactions is 
well known~\cite{Jenk92b,Jenk91a}, and so we will discuss only 
the weak sector.       
In the standard model,  the $\,|\Delta S|=1$,  
$\,|\Delta\bfm{I}|=3/2\,$  weak transitions are described by   
an effective Hamiltonian that transforms as   
$(27_{\rm L}^{},1_{\rm R}^{})$  under chiral rotations.   
At lowest order in  $\chi$PT,  the Lagrangian for such weak   
interactions of baryons that has the required transformation   
properties is~\cite{Abde99b,He98}  
\begin{eqnarray}   \label{loweak}   
\begin{array}{l}   \displaystyle        
{\cal L}^{\rm w}  \;=\;     
\beta_{27}^{}\, 
T_{ij,kl}^{} \left( \xi\bar{B}_v^{} \xi^\dagger \right) _{\!ki}^{} 
\left( \xi B_v^{} \xi^\dagger \right) _{\!lj}^{}   
\vspace{2ex} \\   \displaystyle  
\;\;\;+\;  
\delta_{27}^{}\, 
T_{ij,kl}^{}\; \xi_{kd}^{} \xi_{bi}^\dagger\; 
\xi_{le}^{} \xi_{cj}^\dagger\; 
\bigl( \bar{T}_v^\mu \bigr) _{abc}^{} 
\bigl( T_{v\mu}^{} \bigr) _{ade}^{} 
\vspace{2ex} \\   \displaystyle  
\;\;\;+\;  {\rm h.c.}   \;, 
\end{array}      
\end{eqnarray}      
where  $\beta_{27}^{}$  ($\delta_{27}^{}$)  is the coupling constant
for the octet (decuplet) sector,  $T_{ij,kl}^{}$  is the tensor that 
project out the  $\,|\Delta S|=1$,  $\,|\Delta\bfm{I}|=3/2\,$  
transitions, and further details are given in Ref.~\cite{Abde99b}.

One can now calculate the decay amplitudes.  
In the heavy-baryon approach, the amplitude for   
$\,B\rightarrow B^\prime\pi\,$  can be written as~\cite{Abde99b}   
\begin{eqnarray}        
\begin{array}{l}   \displaystyle        
\ri {\cal M}^{}_{B_{}\rightarrow B_{}'\pi}   \;=\;  
G_{\rm F}^{} m_{\pi}^2\, \times  
\vspace{2ex} \\   \displaystyle    
\hspace*{3em}  
\bar{u}_{B_{}'}^{} \Bigl(   
{\cal A}^{(\rm S)}_{B_{}^{}B_{}'\pi}   
+ 2 k\cdot S_v^{}\, {\cal A}^{(\rm P)}_{B_{}^{}B_{}'\pi} 
\Bigr) u_{B_{}^{}}^{}   \;,  
\end{array}    
\end{eqnarray}    
where  the superscripts refer  to S- and P-wave contributions, 
the $u$'s  are baryon spinors, 
$k$  is the outgoing four-momentum of the pion, and  
$S_v^{}$  is the velocity-dependent spin operator~\cite{Jenk91a}.

At tree level,  ${\cal O}(1)$  in  $\chi$PT,  contributions to 
the amplitudes come from diagrams each with a~weak vertex from 
${\cal L}^{\rm w}$  in~(\ref{loweak}) and, for the P-waves, 
a vertex from the lowest-order strong Lagrangian.  
At next order in $\chi$PT,  there are amplitudes of order  $m_s^{}$, 
the strange-quark mass,  arising both from one-loop diagrams with 
leading-order vertices and from counterterms. 
Currently, there is not enough experimental input to fix 
the counterterms. 
For this reason, we follow the approach that has been used for 
the $\,|\Delta\bfm{I}|=1/2\,$  amplitudes~\cite{Bijn85,Jenk92a} 
and  calculate only nonanalytic terms up to  
${\cal O}(m_s^{}\ln{m_s^{}})$.  
These terms are uniquely determined from the one-loop amplitudes   
because they cannot arise from local counterterm Lagrangians.   
It is possible to do a~complete calculation at next-to-leading order 
and fit all the amplitudes 
(as was done in Ref.~\cite{Bora99} for the $\,|\Delta\bfm{I}|=1/2\,$ 
sector, without explicitly including the decuplet baryons in  
the effective theory),  but then one loses predictive power, 
given the large number of free parameters available. 
Here, we want to limit ourselves to studying the question of whether 
the lowest-order predictions are subject to large higher-order   
corrections.

To compare our theoretical results with experiment, 
we introduce the amplitudes~\cite{Jenk92a}
\begin{eqnarray}   
s  \;=\;  {\cal A}^{\rm (S)}   \;, \hspace{2em}   
p  \;=\;  -|\bfm{k}| {\cal A}^{\rm(P)}   
\end{eqnarray}      
in the rest frame of the decaying baryon.    
From these amplitudes, we can extract for the S-waves 
the $\,|\Delta\bfm{I}|=3/2\,$  components 
\begin{eqnarray}   
\begin{array}{l}   \displaystyle        
S_{3}^{(\Lambda)}  \;=\;  
\ratio{1}{\sqrt{3}} 
\Bigl( \sqrt{2}\, s_{\Lambda\rightarrow n\pi^0}^{}   
      + s_{\Lambda\rightarrow p\pi^-}^{} \Bigr)   \;,    
\vspace{2ex} \\   \displaystyle  
S_{3}^{(\Xi)}  \;=\;  
\ratio{2}{3} \Bigl( \sqrt{2}\, s_{\Xi^0\rightarrow\Lambda\pi^0}^{}  
                   + s_{\Xi^-\rightarrow\Lambda\pi^-}^{} \Bigr)   \;,  
\vspace{2ex} \\   \displaystyle  
S_{3}^{(\Sigma)}  \;=\;  
-\sqrt{\ratio{5}{18}} 
\Bigl( s_{\Sigma^+\rightarrow n\pi^+}^{}  
      - \sqrt{2}\, s_{\Sigma^+\rightarrow p\pi^0}^{}   
\vspace{1ex} \\   \displaystyle  
\hspace*{8em}  
      -\; s_{\Sigma^-\rightarrow n\pi^-}^{} \Bigr)   \;,
\end{array}   
\end{eqnarray}      
and the  $\,|\Delta\bfm{I}|=1/2\,$  components  
(for  $\Lambda$  and  $\Xi$  decays)  
\begin{eqnarray}   
\begin{array}{l}   \displaystyle        
S_{1}^{(\Lambda)}  \;=\;  
\ratio{1}{\sqrt{3}} 
\Bigl( s_{\Lambda\rightarrow n\pi^0}^{}   
      - \sqrt{2}\, s_{\Lambda\rightarrow p\pi^-}^{} \Bigr)   \;,    
\vspace{2ex} \\   \displaystyle  
S_{1}^{(\Xi)}  \;=\;  
\ratio{\sqrt{2}}{3} 
\Bigl( s_{\Xi^0\rightarrow\Lambda\pi^0}^{}  
       - \sqrt{2}\, s_{\Xi^-\rightarrow\Lambda\pi^-}^{} \Bigr)   \;, 
\end{array}      
\end{eqnarray}      
as well as analogous ones for the P-waves. 
We can then compute from data the ratios collected in  
Table~\ref{ratio,isosymmetric}, which show 
the  $\,|\Delta\bfm{I}|=1/2\,$  rule for hyperon decays.  
The experimental values for  $S_3^{}$  and  $P_3^{}$  
are listed in the column labeled ``Experiment'' in  
Table~\ref{results1}.

\begin{table*}
\caption{\label{ratio,isosymmetric}%
Experimental values of ratios of  $\,|\Delta\bfm{I}|=3/2\,$  to  
$\,|\Delta\bfm{I}|=1/2\,$  amplitudes.
}    
\centering   \small   
\begin{tabular*}{1\textwidth}{@{\extracolsep{\fill}}cccccc} 
\hline \hline      
\hspace{2ex} 
$S_3^{(\Lambda)}/S_{1_{}}^{(\Lambda)^{}}$            & 
$S_3^{(\Xi)}/S_1^{(\Xi)}$                            & 
$S_3^{(\Sigma)}/s_{\Sigma^-\rightarrow n\pi^-}^{}$   &  
$P_3^{(\Lambda)}/P_1^{(\Lambda)}$                    &   
$P_3^{(\Xi)}/P_1^{(\Xi)}$                            &   
$P_3^{(\Sigma)}/p_{\Sigma^+\rightarrow n\pi^+}^{}$             
\vspace{0.5ex} \\ \hline \vspace{-2ex} \\
\hspace{2ex} 
$0.026\pm 0.009$   
& $0.042\pm 0.009$   
& $-0.055\pm 0.020$  
& $0.031\pm 0.037$  
& $-0.045\pm 0.047$  
& $-0.059\pm 0.024$  
\vspace{.3ex} \\  
\hline \hline 
\end{tabular*}   
\vskip 1\baselineskip
\end{table*}

To begin discussing our theoretical results,\footnote{In this work, 
we have assumed isospin invariance (massless $u$- and 
$d$-quarks~\cite{Abde99b}).   
With improved data in the future, a more quantitative analysis will 
have to take isospin breaking into account, as it may generate in 
the  $\,|\Delta\bfm{I}|=1/2\,$  amplitudes corrections comparable 
in size to the  $\,|\Delta\bfm{I}|=3/2\,$  amplitudes, especially 
for the~P-waves~\cite{Malt95}.}   
we note that our calculation yields no contributions to the S-wave 
amplitudes  $S_3^{(\Lambda)}$  and  $S_3^{(\Xi)}$,  as shown in 
Table~\ref{results1}.  
This only indicates that the two amplitudes are predicted to be 
smaller than $S_3^{(\Sigma)}$ by about a~factor of three 
because there are nonvanishing contributions from operators that 
occur at the next order,  ${\cal O}(m_s^{}/\Lambda_{\chi\rm SB}^{})$, 
with  $\,\Lambda_{\chi\rm SB}^{}\sim 1\,\rm GeV\,$  
being the scale of chiral-symmetry breaking.
(An example of such operators is considered in Refs.~\cite{Abde99b,He98}.)  
The experimental values of $S_3^{(\Lambda)}$  and 
$S_3^{(\Xi)}$  are seen to support this prediction.

\begin{table*}  
\caption{\label{results1}%
Summary of results for  $\,|\Delta\bfm{I}|=3/2\,$  components of 
the S- and P-wave  amplitudes to  ${\cal O}(m_s^{}\ln{m_s^{}})$.   
We use the parameter values 
$\,\beta_{27}^{}=\delta_{27}^{}=-0.068 \, 
\sqrt{2}\, f_{\!\pi}^{}G_{\rm F}^{} m_{\pi}^2\,$
and a~subtraction scale  $\,\mu = 1\,\rm GeV.\,$}  
\centering   \small   
\begin{tabular*}{1\textwidth}{@{\extracolsep{\fill}}crrrr}    
\hline \hline 
&& \multicolumn{3}{c}{Theory}  
\\ \cline{3-5}    
\hspace{2ex} Amplitude  &  Experiment\hspace{1ex}    &  
\raisebox{-1ex}{Tree\hspace{1ex}}       &  
\raisebox{-1ex}{Octet\hspace{1.5em}}    &   
\raisebox{-1ex}{Decuplet\hspace{2ex}}   
\\   
&&  
\raisebox{-1ex}{${\cal O}(1)$\hspace{0.5ex}}       &   
\raisebox{-1ex}{${\cal O}(m_s^{}\ln{m_s^{}})$}     & 
\raisebox{-1ex}{${\cal O}(m_s^{}\ln{m_s^{}})$}        
\smallskip \\ \hline      
\vspace{-2.5ex} \\  
$S_3^{(\Lambda)}$ &  $-0.047\pm 0.017$    &   0\hspace{1em} &   
\hspace{5ex} 0\hspace{2.5em}   &  
\hspace{4ex} 0\hspace{2.5em}   
\\
$S_3^{(\Xi)}$          & $ 0.088\pm 0.020$      & 0\hspace{1em}    &  
0\hspace{2.5em}        & 0\hspace{2.5em}        
\\   
$S_3^{(\Sigma)}$       & $-0.107\pm 0.038$      & $-$0.107         &  
$-$0.089\hspace{1.5em} & $-$0.084\hspace{1.5em} 
\\  
$P_3^{(\Lambda)}$      & $-0.021\pm 0.025$      & 0.012            &  
0.005\hspace{1.5em}    & $-$0.060\hspace{1.5em} 
\\  
$P_3^{(\Xi)}$          & $ 0.022\pm 0.023$      & $-$0.037         &   
$-$0.024\hspace{1.5em} & 0.065\hspace{1.5em}    
\\  
\vspace{.3ex}   
$P_3^{(\Sigma)}$       & $-0.110\pm 0.045$      & 0.032            &   
0.015\hspace{1.5em}    & $-$0.171\hspace{1em} 
\vspace{.3ex} \\  
\hline \hline 
\end{tabular*}   
\vskip 1\baselineskip
\end{table*}

The other four amplitudes are predicted to be nonzero.   
They depend on the two weak parameters  $\beta_{27}^{}$  and  
$\delta_{27}$  of  ${\cal L}^{\rm w}\,$  
(as well as on parameters from the strong 
Lagrangian, which are already determined),  
with  $\delta_{27}$  appearing only in loop diagrams.
Since we consider only the nonanalytic part of the loop diagrams, 
and since the errors in the measurements of the P-wave amplitudes 
are larger than those in the S-wave amplitudes, we can take the 
point of view that we will extract the value of $\beta_{27}^{}$
by fitting the tree-level $S_3^{(\Sigma)}$ amplitude to experiment, 
and then treat the tree-level P-waves as predictions and the loop 
results as a~measure of the uncertainties of the lowest-order 
predictions.

Thus we obtain 
$\,\beta_{27}^{}=-0.068\,\sqrt{2}\, f_{\!\pi}^{}  
   G_{\rm F}^{} m_{\pi}^2,\,$  
and the resulting P-wave amplitudes are placed in the column labeled
``Tree'' in  Table~\ref{results1}.  
These lowest-order predictions are not impressive, but they 
have the right order of magnitude and differ from the central 
value of the measurements by at most three standard deviations.  
For comparison, in the $\,|\Delta\bfm{I}|=1/2\,$  case   
the tree-level predictions for the P-wave amplitudes are completely 
wrong~\cite{Bijn85,Jenk92a,Abde99a}, differing from the measurements by 
factors of up to 20.

To address the reliability of the leading-order predictions, 
we look at our calculation of the one-loop corrections,    
presented in two columns in  Table~\ref{results1}.  
The numbers in the column marked  ``Octet''  come from  
all loop diagrams that do not have any decuplet-baryon lines,  
with  $\beta_{27}^{}$  being the only weak parameter in the diagrams.  
Contributions of loop diagrams with decuplet baryons depend on one 
additional constant, $\delta_{27}^{}$,  which cannot be fixed 
from experiment as it does not appear in any of the observed 
weak decays of a~decuplet baryon. 
To illustrate the effect of these terms, we choose  
$\,\delta_{27}^{}=\beta_{27}^{},\,$  a~choice consistent with 
dimensional analysis and the normalization of  ${\cal L}^{\rm w}$,  
and collect the results in the column labeled ``Decuplet''.

We can see that some of the loop corrections in  Table~\ref{results1}  
are comparable to or even larger than the lowest-order results 
even though they are expected to be smaller by about a~factor of 
$\,M_K^2/(4\pi f_{\!\pi}^{})^2\approx 0.2.\,$   
These large corrections occur when several different diagrams 
yield contributions that add up constructively, resulting in 
deviations of up to an order of magnitude from the power-counting 
expectation. 
This is an inherent flaw in a~perturbative calculation where 
the expansion parameter is not sufficiently small and there are many 
loop-diagrams involved.      
We can, therefore, say that these numbers are consistent with naive 
expectations.

Although the one-loop corrections are large, they are all much   
smaller than their counterparts in the  $\,|\Delta\bfm{I}|=1/2\,$  
sector, where they can be as large as 30 times the lowest-order  
amplitude~\cite{Abde99a}  in the the P-wave in  
$\,\Lambda\rightarrow p\pi^-.\,$  
In that sector, the loop dominance in the P-waves was due  
to an anomalously small lowest-order prediction arising from 
the cancellation of two nearly identical  terms~\cite{Jenk92a}. 
Such a cancellation does not happen in the  $\,|\Delta\bfm{I}|=3/2\,$ 
case because each of the lowest-order P-waves has only one  
term~\cite{Abde99b}.

In conclusion, we have discussed  $\,|\Delta\bfm{I}|=3/2\,$  
amplitudes for hyperon nonleptonic decays in  $\chi$PT.   
At leading order, these amplitudes are described in terms of  
only one weak parameter.   
We have fixed this parameter from the observed value of    
the S-wave amplitudes in  $\Sigma$  decays.   
Then we have predicted the P-waves and used our one-loop calculation  
to discuss the uncertainties of the lowest-order predictions.   
Our predictions are not contradicted by current data, but current  
experimental errors are too large for a~meaningful conclusion.   
We have shown that the one-loop nonanalytic corrections have   
the relative size expected from naive power-counting.   
The~combined efforts of E871 and KTeV experiments at Fermilab could  
give us improved accuracy in the measurements of some of the decay   
modes that we have discussed and allow a~more quantitative comparison  
of theory and experiment.


\begin{thebibliography}{9}
\bibitem{Bijn85}J.~Bijnens, H.~Sonoda and M.~B.~Wise, Nucl. 
Phys.~B~261 (1985) 185.   

\bibitem{Geor84}
H.~Georgi, Weak Interactions and Modern Particle Theory, 
The Benjamin/Cummings Publishing Company, Menlo Park, 1984;  
J.~F.~Donoghue, E.~Golowich and B.~R.~Holstein, 
Dynamics of the Standard Model, Cambridge University Press, Cambridge, 
1992.

\bibitem{Jenk92a}E.~Jenkins, Nucl. Phys.~B 375~(1992) 561.

\bibitem{Jenk92b}
E.~Jenkins and A.~Manohar, in Effective Field Theories of 
the Standard Model, edited by U.-G. Meissner, World Scientific, 
Singapore, 1992.  

\bibitem{Caro92}C.~Carone and H.~Georgi, Nucl. Phys.~B~375 
(1992) 243.   
   
\bibitem{Spri95}R.~P.~Springer, hep-ph/9508324;   
Phys. Lett.~B~461 (1999) 167.   
   
\bibitem{Abde99a}   
A.~Abd~El-Hady and J.~Tandean, hep-ph/9908498.  
   
\bibitem{Abde99b} 
A.~Abd~El-Hady, J.~Tandean and G.~Valencia, Nucl. Phys.~A~651 (1999) 71.
   
\bibitem{Jenk91a} 
E.~Jenkins and A.~V.~Manohar, Phys. Lett.~B~255 (1991) 558.

\bibitem{Jenk91b} 
E.~Jenkins and A.~Manohar, Phys. Lett.~B~259 (1991) 353.

\bibitem{Caso98} Review of Particle Physics.   
C.~Caso et. al.  Eur. Phys. J.~C~3 (1998) 1.   
    
\bibitem{He98}X.-G.~He and G.~Valencia, Phys. Lett.~B~409 (1997) 469; 
erratum Phys. Lett.~B~418 (1998) 443.
   
\bibitem{Bora99}B.~Borasoy and B.~R.~Holstein, 
Eur. Phys. J.~C~6 (1999) 85.
   
\bibitem{Malt95}K.~Maltman, Phys. Lett.~B~345 (1995) 541; 
E.~S.~Na and B.~R.~Holstein, Phys. Rev.~D~56 (1997) 4404; 

\end{thebibliography}
\end{document}